\documentclass[11pt]{article}
 \ifx\pdfpagewidth\undefined
 \else %% just for pdf output
 \fi
 \usepackage{amsmath} 
 \usepackage{natbib} 
\usepackage[left,displaymath,mathlines]{lineno}
\nolinenumbers
\date{}
\title{A Note On the Use of Fiducial Limits for Control Charts}

\author{\textsf{Mokin Lee} \\
{\it School of Mechanical and Automotive Engineering}  \\
{\it University of Ulsan} \\
{\it Ulsan, South Korea}
\and
\textsf{Chanseok Park} \\
{\it Department of Mathematical Sciences} \\
{\it Clemson University} \\
{\it Clemson, SC 29634, USA.}
}

\begin{document}

\maketitle

\begin{abstract}
Many of the early works in the quality control literature
construct control limits through the use of graphs and tables
as described in \cite{Wortham/Ringer:1972}.
However, the methods used in
this literature are restricted to using only the values that the
graphs and tables can provide and to the case where the parameters 
of the underlying distribution are known.  In this note, we briefly
describe a technique which can be used to calculate exact control limits
without the use of graphs or tables.
We also describe what are commonly  referred to in the literature as fiducial
limits. Fiducial limits are often used as the limits in control charting
when the parameters of the underlying distribution are unknown.
\end{abstract}

\clearpage
%%===============================================
\section{Introduction}
%%===============================================
In many applications using attribute (count) data,
the number of nonconformities often has discrete distributions such  as
the binomial, Poisson and geometric. In this note, we provide the fiducial
limits for these distributions because, in certain cases,  these limits
are often used as the upper and lower  limits  for control charts.

Early works related to obtaining exact control limits
were based on the use of graphs and tables. 
For example, \cite{Wortham/Ringer:1972} provide a good review 
on exact control limits using graphs and tables
when the true parameter of the underlying distribution is known. 
These graphs and tables were
used to approximate or interpolate the values needed to obtain an interval 
with $1-\alpha$ confidence level when the parameters of the underlying distribution
were known. 
Before the advent of powerful
computers and software, it was often troublesome and time-consuming to
compute exact control limits.  Thus, the methods that were based on the
use of tables and graphs were prevalent in the literature because of
their simplicity and convenience.

Unfortunately, the studies that use these graphs
and tables are restricted to using the specific confidence interval
values that can be inferred from these tables and graphs. 
Additionally, errors are difficult to avoid because of the required use of visual
interpolation. 
However, with powerful computers and accessible software available, 
it has now become quite trivial to calculate exact
control limits without the use of graphs and tables. 
Exact control limits are obtained by calculating the corresponding quantiles of
the underlying distribution involved. 
These days, there are many statistical software programs that provide the quantiles for any
reasonably common distribution. 
For example, in \cite{R}, a non-commercial
open source statistical software package for statistical computing and
graphics, the binomial and Poisson quantile functions are given by
\texttt{qbinom(~)} and \texttt{qpois(~)} respectively. 
R can be obtained at no cost from \texttt{http://www.r-project.org}.
Other commercial statistical software such
as Minitab and SAS also provide the quantile functions for the
commonly known distributions.

Another shortcoming of the control limit methods
that use the graphs and tables in \cite{Wortham/Ringer:1972} is that,
in order to use these tables, one has to assume that the parameter of
the underlying distribution is known. 
In many practical examples, the true parameters are not known. 
As aforementioned, if the true parameter is known, the
exact limits are easily calculated by using the quantile functions in
the various statistical software packages available. If the parameters
are unknown, then the control limits can be obtained by inverting the
relation between the tails and the parameters. These control limits are
then referred to as fiducial limits in the statistics literature and
are explained in more detail in the following section.

%============================
\section{Fiducial Limits for the discrete distributions}
%============================
Many authors, including 
\cite{Clopper/Pearson:1934}, \cite{Garwood:1936}, \cite{Stevens:1950}, 
\cite{Bickel/Kjell:1977}, 
\cite{Kendall/Stuart:1979},
mention the exact or fiducial confidence limits 
for several discrete distributions, 
along with the classical confidence limits based on the central limit theorem. 
In this note, we provide a review of several fiducial limits which can be useful 
for various control charting methods including the $p$, $u$, $c$ and $g$ charts. 

%----------------------------
\subsection{The binomial distribution}
%----------------------------
Let $X$ denote a binomial random variable with size $n$ and Bernoulli probability $p$.
The fiducial limits for $p$ are obtained by inverting the following two equal tails for $p$
%--------
\begin{linenomath}
\begin{align}
\sum_{i=0}^{x} \binom{n}{i} p^i (1-p)^{n-i} &= \frac{\alpha}{2} \\
\intertext{and}
\sum_{i=x}^{n} \binom{n}{i} p^i (1-p)^{n-i} &= \frac{\alpha}{2}, 
\end{align}
\end{linenomath}
%--------
where $x=1,2,\ldots, n-1$.
The lower and upper limits with $1-\alpha$ confidence level are given by 
%--------
\begin{linenomath}
\begin{align}
p_L = \Bigg\{ 1 + \frac{n-x+1}{x \cdot F_{2x,2(n-x+1)}^{(\alpha/2)}} \Bigg\}^{-1}
\intertext{and}
p_U = \Bigg\{ 1 + \frac{n-x}{ (x+1) \cdot F_{2(x+1),2(n-x)}^{(1-\alpha/2)}} \Bigg\}^{-1} ,
\end{align}
\end{linenomath}
%--------
where $F_{\nu_1,\nu_2}^{(\xi)}$ denotes the $\xi$ quantile of the 
$F$ distribution with degrees of freedom $\nu_1$ and $\nu_2$.
Here, the lower limit is $0$ if $x=0$ and the upper limit is $1$ if $x=n$. 
Derivations of the fiducial limits for $p$ based on the $F$ distribution are given by 
\cite{Blyth:1986}, 
\cite{Hald:1952}, 
and \cite{Leemis/Trivedi:1996}.
It should be noted that the quantile of the $F$ distribution is calculated by using 
\texttt{qf(~)} in R.

%----------------------------
\subsection{The Poisson distribution}
%----------------------------
\cite{Garwood:1936} provide the fiducial limits for the Poisson distribution.
Let $X_1, X_2, \ldots, X_n$ be a random sample from a Poisson distribution with mean $\lambda$.
For convenience, let $Y = \sum_{j=1}^{n} X_{j}$. 
Then, the random variable $Y$ has a Poisson distribution with mean $n \lambda$.
Let $Y=y$, then 
the lower and upper limits with $1-\alpha$ confidence level are given by 
%--------
\begin{linenomath}
\begin{align}
\lambda_L = \frac{1}{2n} \Gamma_{y,2}^{(\alpha/2)} 
\intertext{and}
\lambda_U = \frac{1}{2n} \Gamma_{y+1,2}^{(1-\alpha/2)}, 
\end{align}
\end{linenomath}
%--------
where $\Gamma_{a,b}^{(\xi)}$ is the $\xi$ quantile of the gamma distribution 
with parameters $a$ and $b$. Here the lower limit is $0$ if $y=0$. 
It should be noted that the quantile of the gamma distribution is calculated by using 
\texttt{qgamma(~)} in R.

%----------------------------
\subsection{The geometric distribution}
%----------------------------
Let $X_1, X_2, \ldots, X_n$ be a random sample from a geometric distribution with 
Bernoulli parameter $p$.
For convenience, let $Y = \sum_{j=1}^{n} X_{j}$. 
Then, the random variable $Y$ has a negative binomial distribution with size $n$ and 
Bernoulli parameter $p$. 
Let $Y=y$, then 
the lower and upper limits with $1-\alpha$ confidence level equal
%--------
\begin{linenomath}
\begin{align}
p_L = \Bigg\{ 1 + { \frac{y+1}{n} \cdot F_{2(y+1),2n}^{(1-\alpha/2)}} \Bigg\}^{-1}
\intertext{and}
p_U = \Bigg\{ \frac{ {n} \cdot F_{2n,2y}^{(1-\alpha/2)} }
              { y +  {n} \cdot F_{2n,2y}^{(1-\alpha/2)}} \Bigg\}^{-1} ,
\end{align}
\end{linenomath}
%--------
where the upper limit is $1$ if $y=0$.
These limits are provided in Exercise~9.22 of \cite{Casella/Berger:2002}.

%----------------------------
\section{Illustrative Examples}
%----------------------------

%%==================================================================
%% \bibliographystyle{unsrt}
%% \bibliographystyle{apalike}
%% \bibliographystyle{natbib}
%%\bibliography{/home/cspark/library/TeX/bib/STAT,/home/cspark/library/TeX/bib/ENG}

%%==================================================================

%%===============================================
\end{document}